\documentclass[12pt]{article}
\usepackage{amsmath,epsfig}
\usepackage{amsmath}
\usepackage{amssymb}
\usepackage{mathrsfs}
\usepackage{amsmath,epsfig}
\usepackage{graphicx}
\usepackage{amsmath,epsfig}
\usepackage{epsfig}
\newsavebox{\PSLASH}
\sbox{\PSLASH}{$p$\hspace{-1.8mm}/}

\begin{document}

\vspace{4cm}
\begin{center}
{\Large\bf{Constraints on the Masses of Fourth Generation Quarks }}\\
\vspace{1cm}
{\bf S. Hosseini$^{\dagger}$, M. Mohammadi Najafabadi$^{\ddagger,}$\footnote{\normalsize{Corresponding author email address:
mojtaba@ipm.ir}}, A. Moshaii$^{\dagger,\ddagger}$, Y. Radkhorrami$^{\dagger}$, N. Tazik$^{\S}$}\\
\vspace{0.5cm}
{\sl ${\ ^{\ddagger}}$ Physics Department, Tarbiat Modarres University, Tehran, Iran}\\
{\sl ${\ ^{\ddagger}}$  School of Particles and Accelerators, 
Institute for Research in Fundamental Sciences (IPM) \\
P.O. Box 19395-5531, Tehran, Iran}\\
{\sl ${\ ^{\S}}$ Physics Department, Semnan University, Semnan, Iran}\\
\vspace{3cm}
 \textbf{Abstract}\\
 \end{center}

We study the one loop contribution of the down type quark of the
SM-like fourth generation ($b^{\prime}$) on the top quark electric
dipole moment. Using the known limits on the top quark electric
dipole moments (EDM), we place limits on the $b^{\prime}$ mass.
Then from the estimated ratio for the masses of the fourth
generation of quarks from other studies and the achieved bound
from top quark EDM on $m_{b^{\prime}}$, we obtain a limit for
the up type quark of fourth generation ($t^{\prime}$) mass.

\newpage

\section{Introduction}

The Standard Model (SM) of particle physics is in a very good
agreement with present experimental data. Nonetheless, it is
believed to leave many questions unanswered, and this belief has resulted in numerous
theoretical and experimental attempts to discover a more
fundamental underlying theory. Various types of experiments may
expose the existence of physics beyond the SM, including the
search for direct production of exotic particles at high energy
colliders. A complementary approach in hunting for new physics is
to examine its indirect effects in higher order processes.

As mentioned, the SM with three generations of quarks and leptons
is in excellent agreement with the current experimental data.
However, the SM does not explain the fermion mass hierarchy and
it also is not able to explain why there are precisely three
families. Several models have been proposed to solve the
shortcomings of the SM through the introduction of new
generations of quarks and leptons. While some models beyond SM,
such as Grand Unified Theories (GUT), predict the new generations
of quarks or leptons. The strong CP problem is solvable by
requiring additional quarks. The weak CP violation may be
accommodated through the KM mechanism, while the strong CP issue
can be solved in a model with two additional flavor of quarks by
spontaneous CP violation. Another motivation for fourth generation
is that in the literature it has been shown that a
non-supersymmetric model with four generations can have
successful unification of gauge couplings at the unification
scale. In the scenarios of gauge mediated supersymmetry breaking
additional generations of quarks and leptons arise automatically.
More details can be found in \cite{frampton1},\cite{frampton2},
\cite{barger} and references therein. It should be mentioned that
the fourth generation of quarks and leptons can be a chiral
doublet (SM-like fermion generation) or non-chiral doublet (also
known as vector-like). For example, in Grand Unified Theories all
types of additional fermions are possible. While in models which
attempt to solve the strong CP problem only non-chiral doublet of
quarks are considered. There have been already many indirect and
direct studies on the fourth generation of quarks. For example,
the effects of a vector-like fourth generation of quarks on the
width of $Z$ boson and forward-backward asymmetry has been
studied in \cite{tadashi}. Bounds on the mixing of the SM down
type quarks with new vector-like singlet quarks derived in
\cite{silva}. In \cite{bogdan}, the pair production of
$t^{\prime}$ quarks at Tevatron has been studied. It has been
shown that the production cross section for
$t^{\prime}\bar{t^{\prime}}$ at hadron colliders could be
considerably higher than QCD prediction if a gluon-prime (a
massive color octet vector boson) is present in the theory. In
\cite{cdf}, using a data sample corresponding to 2.8 fb$^{-1}$ of
integrated luminosity recorded by CDF experiment in proton
anti-proton collisions, a limit was set on the production cross
section of $t^{\prime}\bar{t^{\prime}}$. From this limit a lower
limit of 311 GeV/c$^{2}$ was derived for a new heavy top-like
quark.

Since top quark is far more massive than other SM fermions, its
interactions may be quite sensitive to new physics originating at
higher scale \cite{beneke},\cite{fcnc},\cite{wp}. Hence, the study of interactions of
fourth generation of quarks with top quark might give interesting results
about the fourth generation.

In this article, our aim is to constraint the mass of
the down type quark of the fourth generation ($b^{\prime}$) using the
one loop contribution of the $b^{\prime}$ in the electric dipole moment (EDM) of top quark.
In the analysis, we will use the estimated bounds on the
EDM's of top quark to constraint the mass of $b^{\prime}$.
In \cite{frampton2}, it has been shown that for the chiral doublet
of $(t^{\prime},b^{\prime})$ the ratio of masses is 1.1 or less 
($m_{t^{\prime}}/m_{b^{\prime}}\leq 1.1$).
Using this value, the bound on $m_{t^{\prime}}$ is also estimated.

\section{The Contribution of the $b^{\prime}$ in the Top Quark EDM}
In this work, we examine
the properties of masses of fourth generation of quarks.
Similar to the interaction of $Wtb$, a general effective Lagrangian
for the interaction of $Wtb^{\prime}$ can be written in the following form:
\begin{eqnarray}\label{lag}
{\cal L}_{Wtb^{\prime}} = \frac{g}{\sqrt{2}}\bar{t}\gamma^{\mu}
\left( g_{L}P_{L} + g_{R}P_{R}\right)b^{\prime}W_{\mu}
\end{eqnarray}
where $P_{L}(P_{R})$ are the left-handed (right-handed) projection operators. The
$g_{L},g_{R}$ coefficients are complex in general. This signifies the CP violating
effects. These coefficients include the mixing factor between the fourth
family and the top quark ($V_{tb^{\prime}}$) in the generalized CKM matrix.
For the interaction of $Wtb$, these factors have been estimated from
different studies. For example, from the B decay processes the limits on $g_{L},g_{R}$ are:
$Re(g_{R})\leq 4\times 10^{-3}$, $Im(g_{R})\leq 10^{-3}$ and
$Im(g_{L})\leq 3 \times 10^{-2}$ \cite{b1},\cite{b2},\cite{b3}.

\begin{figure}
\centering
\includegraphics[width=12cm,height=6cm]{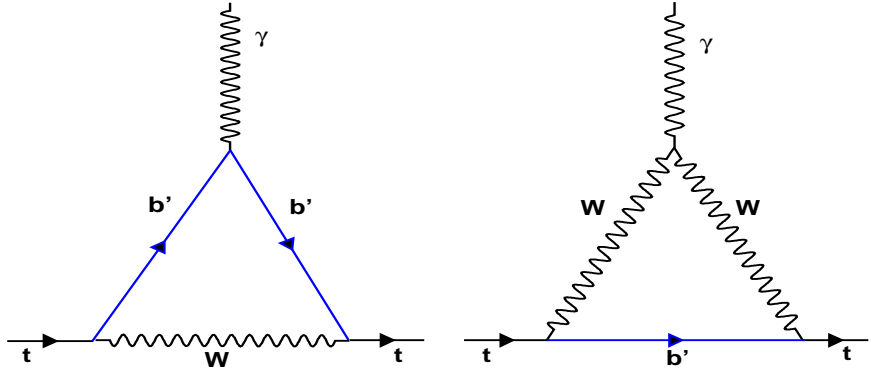}
\caption{\label{vertex}Feynman diagrams contributing to the on shell
$\bar{t}t\gamma$ vertex.}
\end{figure}

The introduced Lagrangian in Eq.\ref{lag} induces an electric
dipole moment for the top quark at the one loop level via the
Feynamn diagrams shown in Fig.\ref{vertex}. After calculation of the one loop
corrections to the vertex of $\bar{t}t\gamma$ shown in Fig.\ref{vertex},
we find some terms with different structures. The coefficient of
the structure of $\sigma_{\mu\nu}\gamma_{5}q^{\nu}$ gives the top quark electric
dipole moment where $q^{\nu}$ is the four momentum of photon \cite{edm1}.
It should be mentioned that this structure arises via radiative
corrections and does not exist at tree level.
After all calculation, the top EDM is found as follows:
\begin{equation}
\label{topEDM} {\text Re}(d_{top})=-\frac{e}{m_{W}}\frac{3\,\alpha}{32
\pi}\,\frac{m_{b^{\prime}}}{m_{W}}\,\left(V_{1}(x_{b^{\prime}},x_{W})+
\frac{1}{3}\,V_{2}(x_{b^{\prime}},x_{W})\right)\,{\text Im}\left(g_{L} g^{*}_{R}\right),
\end{equation}
\noindent where $x_{a}=m_a^{2}/m_t^{2}$. The
$V_{1,2}$ are the functions stand for the contribution of the Feynman
diagram where the photon emerges from the $W$ boson and the $b^{\prime}$
quark line, respectively. They have the following forms:
\begin{eqnarray}
V_{1}=-\left(4x_W-x_{b^{\prime}}+1\right)f(x_{b^{\prime}},x_W)-
\left(x_{b^{\prime}}^2+4x_W^2-5x_{b^{\prime}}x_W-3x_W-2x_{b^{\prime}}+1\right)g(x_{b^{\prime}},x_W)\nonumber \\
V_{2}=-\left(4x_W-x_{b^{\prime}}+1\right)f(x_W,x_{b^{\prime}})+
\left(x^{2}_{b^{\prime}}+4x_W^2-5x_{b^{\prime}}x_W-3x_W-2x_{b^{\prime}}+1\right)g(x_W,x_{b^{\prime}})
\end{eqnarray}
\noindent where the functions of $f$ and $g$ are as follows:
\begin{eqnarray}
f(a,b)&=&\left(\frac{1+a-b}{2}\right)\log\left(\frac{b}{a}\right)+\sqrt{(1-a-b)^2-4ab}\,\times{\rm
ArcSech}\left(\frac{2 \sqrt{ab}}{a+b-1}\right)+2 \nonumber \\
g(a,b)&=&-\frac{1}{2}\log\left(\frac{b}{a}\right)-\frac{1+a-b}{\sqrt{(1-a-b)^2-4ab}}\,\times{\rm
ArcSech}\left(\frac{2 \sqrt{ab}}{a+b-1} \right) \nonumber
\end{eqnarray}

\section{The Limits on $m_{t^{\prime}},m_{b^{\prime}}$}

\begin{figure}
\centering
\includegraphics[width=12cm,height=6cm]{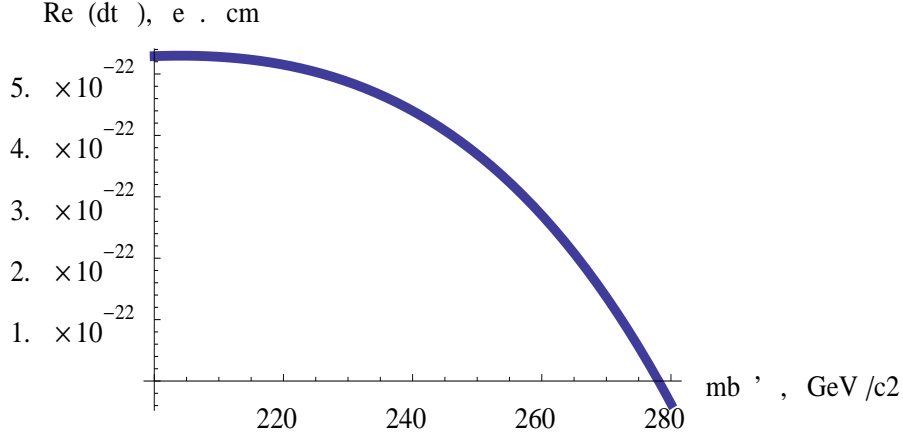}
\caption{The real part of top electric dipole moment versus $m_{b^{\prime}}$
 when ${\text Im}\left(g_{L} g^{*}_{R}\right) = 10^{-3}$.}
\label{dtmb}
\end{figure}

In \cite{toscano2}, the authors have predicted an upper bound for
the top quark EDM. In that paper, a source of CP violation mediated 
by $WW\gamma$ vertex has been analyzed using the effective Lagrangian
technique and its implications on the CP-odd electromagnetic properties 
of the Standard Model particles have been studied. The contribution of 
$WW\gamma$ vertex to the EDM of charged leptons and quarks has been 
calculated.
Their estimate for the top quark EDM is
$1.6\times 10^{-22}$ e.cm. In Eq.\ref{topEDM}, a reasonable
assumption is to set $Im\left(g_{L} g^{*}_{R}\right)$ to a value
around $10^{-3}$ \cite{hou}. Under this assumption and by using the bound of
the top EDM, the lower limit of $268$ GeV/c$^{2}$ is achieved for
the mass of the down type quark in the fourth family of quarks.
Fig.\ref{dtmb} presents the dependency of the top quark electric 
dipole moment on the $m_{b^{\prime}}$ for $Im\left(g_{L} g^{*}_{R}\right) = 10^{-3}$.

Obviously, this lower limit depends on the quantity of
$Im\left(g_{L} g^{*}_{R}\right)$. However, it is not highly
dependent on $Im\left(g_{L} g^{*}_{R}\right)$. When $Im\left(g_{L}
g^{*}_{R}\right)$ changes from $10^{-3}$ to $1$, the lower limit
on the $m_{b^{\prime}}$ varies only up to $3\%$. One of the recent
estimated lower bound on the $m_{b^{\prime}}$ is 199 GeV/c$^{2}$
\cite{pdg}. Hence, the constraint obtained in the current study
is compatible with other studies and the lower bound on the
$m_{b^{\prime}}$ is slightly increased.

For the chiral doublet of $(t^{\prime},b^{\prime})$, the
electroweak precision measurements predict that
$\frac{m_{t^{\prime}}}{m_{b^{\prime}}} \leq 1.1$. Combining this
with the bound obtained from top EDM on the mass of $b^{\prime}$
immediately gives the lower limit of 294.8 GeV/c$^{2}$ on the mass
of $t^{\prime}$. This value also confirms the achieved constraint
from other studies (311 GeV/c$^{2}$) which mentioned in the
introduction \cite{cdf}.

\section{Conclusion}

In this paper, we tried to extract a limit on the mass of the
down type quark of the SM-like fourth generation ($b^{\prime}$) by
employing a general Lagrangian for the interaction of
$W-t-b^{\prime}$ with complex left-handed and right-handed
couplings. This general Lagrangian produces an electric dipole
moment for the top quark at one loop level which contains
$m_{b^{\prime}}$. From the estimated upper limit on the top quark
EDM, a lower limit of $268$ GeV/c$^{2}$ predicted for the mass of
$b^{\prime}$. From electroweak precision data, other studies have
been shown that for the chiral doublet of
$(t^{\prime},b^{\prime})$ the ratio of masses is 1.1 or less. It
turns out that the lower bound on the mass of $t^{\prime}$ is
294.8 GeV/c$^{2}$. These results are regular and compatible with
those obtained from other studies and the bound on $m_{b^{\prime}}$ is 
slightly increased.

\end{document}